\documentclass[12pt]{article}
\usepackage{paper2e}
\usepackage{mydefs2e}

\begin{document}
\begin{titlepage}

\preprint{UMD-PP-97-43\\
{\tt hep-ph/9611343}\\
November, 1996}

\title{A Supersymmetric Composite Model\\\medskip
of Quarks and Leptons}

\author{Markus A.~Luty%
\footnote{\tt mluty@physics.umd.edu}
and Rabindra N.~Mohapatra%
\footnote{\tt rmohapatra@umdhep.umd.edu}}

\address{Department of Physics\\
University of Maryland\\
College Park, Maryland 20742}

\begin{abstract}
We present a class of supersymmetric models with complete
generations of composite quarks and leptons using recent
non-perturbative results for the low energy dynamics of
supersymmetric QCD.
In these models, the quarks arise as composite ``mesons'' and 
the leptons emerge as composite ``baryons.''
The quark and lepton flavor symmetries are linked at the preon level.
Baryon number violation is automatically suppressed by accidental 
symmetries.
We give some speculations on how this model might be made realistic.
\end{abstract}

\end{titlepage}

% ----------------------------------------------------------------------------
% Start of main Text
% ----------------------------------------------------------------------------
\section{Introduction}
There has been a revival of interest in the old idea that quarks
and leptons may be composite \cite{OldComposite,CompositeReview} due
to recent progress in the understanding of the non-perturbative
low energy dynamics of ${\cal N} = 1$ supersymmetric gauge theories
\cite{SeibergVacua,MoreSeiberg}.
In special cases such as $\SU{N}$ gauge theory with $N + 1$ flavors,
it is believed that the low-energy 
degrees of freedom are composite gauge-invariant chiral superfields
interacting via a dynamical superpotential \cite{SeibergVacua}.
While this picture of the dynamics cannot be rigorously proven, it 
passes an impressive number of consistency checks, far beyond the 
't~Hooft anomaly matching conditions that are the main dynamical input 
in traditional composite models.
For example, the models of \Ref{SeibergVacua} have moduli spaces of
supersymmetric vacua, and the dynamics interpolates consistently
between vacua in which the theory is strongly and weakly coupled.
Also, by adding masses for some of the fields in the model, one can 
relate the dynamics in these models to models with fewer flavors, 
which are well understood \cite{ADS}.
This gives us confidence in the dynamical picture 
advocated in \Ref{SeibergVacua}, and holds out the exciting 
possibility that one can construct composite models without
\emph{ad hoc} dynamical assumptions.

By gauging a subgroup of the global symmetry group of a 
strongly-coupled supersymmetric gauge theory with the standard model
gauge group, one can find models with composite chiral superfields 
with the quantum numbers of the standard model fields.
Such models generally do not break supersymmetry, and have unwanted 
particles and symmetries that must be removed by adding more fields 
and interactions to obtain a realistic model.
Such a strategy has been successfully pursued in several recent papers
by Nelson and Strassler \cite{NelsonStrassler}.
In their models, the right-handed down quarks $\bar{D}$ and the lepton
doublets $L$ are elementary whereas the remaining quark and lepton
fields are composite.

In this note, we present a class of models where all standard
model quarks and leptons emerge as bound states from the confining 
dynamics of an $\SU{N}$ gauge theory.
These models display an interesting unification between the quarks and
leptons at the preon level, in the sense that quarks and leptons are 
composites of the same preons.
In a multi-generation version of these models, the same preonic horizontal
symmetries act on up-type (respectively down-type) quarks and 
charged leptons (respectively right-handed neutrinos).
Because the quarks correspond to higher-dimension operators in the 
preonic theory, the lowest-dimension baryon number violating operators
are dimension 7 in this model, and baryon number violation is highly 
suppressed.

The resulting models are far from realistic, and we only make some brief 
remarks on extending the models to make them more realistic.
(For one application, see Ref.~\cite{LowSUSY}.)
We hope that some of the ingredients of these models will be relevant 
for the understanding of the flavor problem or other aspects of 
quark--lepton physics.

% ----------------------------------------------------------------------------
\section{One Composite Generation}
% ----------------------------------------------------------------------------
We first illustrate the basic ideas with a model
for one generation of composite quarks and leptons.
We describe the model by giving the representations of the fields 
under the group
\beq
\SU{4}_{H} \times \SU{3}_{C} \times \SU{2}_{W} \times \U1_{Y}
\times \bigl[ \U1_{B} \times \U1_{L} \bigr],
\eeq
where $\SU{4}_{H}$ is a strong ``hypercolor'' gauge group,
$\SU{3}_{C} \times \SU{2}_{W} \times \U1_{Y}$ is the usual standard 
model gauge group, and $\U1_{B} \times \U1_{L}$ are the global
baryon and lepton number symmetries.
We explicitly display the baryon and lepton number quantum numbers to 
show that baryon number and lepton number are anomaly-free conserved
quantum numbers.
The theory consists of the preon fields
\beq\bal
P_{Q} &\sim (\Yfund, 1, \Yfund, 0)
\times (\sfrac{1}{4}, \sfrac{1}{4}),
\\
\bar{P}_{U} &\sim (\bar{\Yfund}, 1, 1, -1)
\times (-\sfrac{1}{4}, -\sfrac{1}{4}),
\\
\bar{P}_{D} &\sim (\bar{\Yfund}, 1, 1, 1)
\times (-\sfrac{1}{4}, -\sfrac{1}{4}),
\\
P_{C} &\sim (\Yfund, \bar{\Yfund}, 1, -\sfrac{1}{3})
\times (-\sfrac{1}{12},\, \sfrac{1}{4}),
\\
\bar{P}_{C} &\sim (\bar{\Yfund}, \Yfund, 1, \sfrac{1}{3})
\times (\sfrac{1}{12}, -\sfrac{1}{4}),
\eal\eeq
where $\Yfund$ ($\bar{\Yfund}$) denote the fundamental 
(antifundamental) \rep of the relevant $\SU{N}$ group.
If we combine $\bar{P}_{U,D}$ into an $\SU{2}_{R}$ doublet, we see 
that the hypercharge assignments obey $Y = T_{3R} + (B - L)$, so this 
model can be embedded in a left-right symmetric model.
If we ignore the weak standard model gauge couplings, this theory is
$\SU{4}$ \susc QCD with 5 flavors.
The analysis of \Ref{SeibergVacua} implies that the low energy degrees
of freedom in this model have quantum numbers follows.
There are composite ``meson'' states
\beq\bal
Q &\sim \bar{P}_{C} P_{Q} \sim
(1, \Yfund, \Yfund, \sfrac{1}{3})
\times (\sfrac{1}{3}, 0),
\\
\bar{U} &\sim \bar{P}_{U} P_{C} \sim
(1, \bar{\Yfund}, 1, -\sfrac{4}{3})
\times (-\sfrac{1}{3}, 0),
\\
\bar{D} &\sim \bar{P}_{D} P_{C} \sim
(1, \bar{\Yfund}, 1, \sfrac{2}{3})
\times (-\sfrac{1}{3}, 0),
\\
\bar{\Phi}_{U} &\sim \bar{P}_{U} P_{Q} \sim
(1, 1, \Yfund, -1)
\times (0, 0),
\\
\bar{\Phi}_{D} &\sim \bar{P}_{D} P_{Q} \sim
(1, 1, \Yfund, 1)
\times (0, 0),
\\
A &\sim (\bar{P}_{C} P_{C})_{8} \sim
(1, \hbox{\bf Ad}, 1, 0)
\times (0, 0),
\\
T &\sim (\bar{P}_{C} P_{C})_{1} \sim
(1, 1, 1, 0)
\times (0, 0),
\eal\eeq
and composite ``baryons''
\beq\bal
L &\sim P_{C}^{3} P_{Q} \sim
(1, 1, \Yfund, -1)
\times (0, 1),
\\
\bar{E} &\sim \bar{P}_{C}^{3} \bar{P}_{U} \sim
(1, 1, 1, 2)
\times (0, -1),
\\
\bar{N} &\sim \bar{P}_{C}^{3} \bar{P}_{D} \sim
(1, 1, 1, 0)
\times (0, -1),
\\
X &\sim P_{C}^{2} P_{Q}^{2}
\sim (1, \Yfund, 1, -\sfrac{2}{3})
\times (\sfrac{1}{3}, 1),
\\
\bar{X} &\sim \bar{P}_{C}^{2} \bar{P}_{U} \bar{P}_{D}
\sim (1, \bar{\Yfund}, 1, \sfrac{2}{3})
\times (-\sfrac{1}{3}, -1).
\eal\eeq
We see that the composite fields include one generation of quarks
and leptons ($Q$, $\bar{U}$, $\bar{D}$, $L$, $\bar{E}$, plus
right-handed neutrinos $\bar{N}$) and several additional fields:
$\bar{\Phi}_{U,D}$ are Higgs doublets,
$A$ is a color octet,
$T$ is a singlet,
and $X, \bar{X}$ are leptoquarks.
The theory has a dynamical \spot
\beq[Wdyn]
\bal
W_{\rm dyn} &= L \bar{\Phi}_{U} \bar{E}
+ L \bar{\Phi}_{D} \bar{N}
+ L Q \bar{X}
+ X \bar{U} \bar{E}
+ X \bar{D} \bar{N}
+ X A \bar{X} + X T \bar{X}
\\
& \quad - \hbox{determinant},
\eal\eeq
where ``determinant'' denotes terms proportional to 5 powers of 
composite meson fields.

This is a promising starting point for constructing a realistic model
of composite quarks and leptons.
We find it striking that the composite particles with standard model
quantum numbers (including lepton  and baryon numbers) emerge from 
such a simple model.
In particular, the leptons are ``unified'' with the quarks (in the 
sense that they are composites of the same preons) without an 
$\SU{4}$ Pati--Salam symmetry.

% ----------------------------------------------------------------------------
\section{Three Composite Generations}
% ----------------------------------------------------------------------------
We can extend the model above to three generations in at least two ways.
The simplest is to copy the above structure three times, resulting in 
a model with hypercolor group
$\SU{4} \times \SU{4} \times \SU{4}$.
A more interesting way is based on an $\SU{8}_{H}$ hypercolor group with 9 
flavors, where part of the additional global symmetry is interpreted as 
a ``horizontal'' flavor symmetry.

We describe the second model by giving the representations of the fields 
under the group
\beq\bal
\SU{8}_{H} &\times \SU{3}_{C} \times \SU{2}_{W} \times \U1_{Y}
\\
&\times \bigl[ \SU{3}_{Q} \times \SU{3}_{U} \times
\SU{3}_{D} \times \U1_{B} \times \U1_{L} \bigr],
\eal\eeq
where the groups in brackets are global symmetries, some of which
will have to be broken in a realistic model.
The preon fields are
\beq\bal
P_{Q} &\sim (\Yfund, 1, \Yfund, 0)
\times (\Yfund, 1, 1, \sfrac{1}{8}, \sfrac{1}{8}),
\\
\bar{P}_{U} &\sim (\bar{\Yfund}, 1, 1, -1)
\times (1, \bar{\Yfund}, 1, -\sfrac{1}{8}, -\sfrac{1}{8}),
\\
\bar{P}_{D} &\sim (\bar{\Yfund}, 1, 1, 1)
\times (1, 1, \bar{\Yfund}, -\sfrac{1}{8}, -\sfrac{1}{8}),
\\
P_{C} &\sim (\Yfund, \bar{\Yfund}, 1, -\sfrac{1}{3})
\times (1, 1, 1, -\sfrac{5}{24}, \sfrac{1}{8}),
\\
\bar{P}_{C} &\sim (\bar{\Yfund}, \Yfund, 1, \sfrac{1}{3})
\times (1, 1, 1, \sfrac{5}{24}, -\sfrac{1}{8}).
\eal\eeq
This theory has confining dynamics similar to the model in the 
previous section.
The composite fields in this model consist of the ``meson'' fields
\beq\bal
Q &\sim \bar{P}_{C} P_{Q} \sim (1, \Yfund, \Yfund, \sfrac{1}{3})
\times (\Yfund, 1, 1, \sfrac{1}{3}, 0),
\\
\bar{U} &\sim \bar{P}_{U} P_{C} \sim
(1, \bar{\Yfund}, 1, -\sfrac{4}{3})
\times (1, \bar{\Yfund}, 1, -\sfrac{1}{3}, 0),
\\
\bar{D} &\sim \bar{P}_{D} P_{C} \sim
(1, \bar{\Yfund}, 1, \sfrac{2}{3})
\times (1, 1, \bar{\Yfund}, -\sfrac{1}{3}, 0),
\\
\bar{\Phi}_{U} &\sim \bar{P}_{U} P_{Q} \sim (1, 1, \Yfund, -1)
\times (\Yfund, \bar{\Yfund}, 1, 0, 0),
\\
\bar{\Phi}_{D} &\sim \bar{P}_{D} P_{Q} \sim (1, 1, \Yfund, 1)
\times (\Yfund, 1, \bar{\Yfund}, 0, 0),
\\
A &\sim (\bar{P}_{C} P_{C})_{8} \sim (1, \hbox{\bf Ad}, 1, 0)
\times (1, 1, 1, 0, 0),
\\
T &\sim (\bar{P}_{C} P_{C})_{1} \sim (1, 1, 1, 0)
\times (1, 1, 1, 0, 0),
\eal\eeq
and ``baryon'' fields
\beq\bal
L &\sim P_{C}^{3} P_{Q}^{5} \sim (1, 1, \Yfund, -1)
\times (\bar{\Yfund}, 1, 1, 0, 1)
\\
\bar{E} &\sim \bar{P}_{C}^{3} \bar{P}_{U}^{2} \bar{P}_{D}^{3}
\sim (1, 1, 1, 2)
\times (1, \Yfund, 1, 0, -1),
\\
\bar{N} &\sim \bar{P}_{C}^{3} \bar{P}_{U}^{3} \bar{P}_{D}^{2}
\sim (1, 1, 1, 0)
\times (1, 1, \Yfund, 0, -1),
\\
X &\sim P_{C}^{2} P_{Q}^{6}
\sim (1, \Yfund, 1, -\sfrac{2}{3})
\times (1, 1, 1, 1, 1)
\\
\bar{X} &\sim \bar{P}_{C}^{2} \bar{P}_{U}^{3} \bar{P}_{D}^{3}
\sim (1, \bar{\Yfund}, 1, \sfrac{2}{3})
\times (1, 1, 1, -1, -1).
\eal\eeq
This model produces precisely three generations of quarks and leptons, 
plus right-handed neutrinos.
The charged leptons transform under the $\SU{3}_{U}$ horizontal
symmetry,  while the right-handed neutrinos transform under
$\SU{3}_{D}$.
Note also that the Higgs multiplets $\bar{\Phi}_{U,D}$ transform under the 
horizontal symmetries, while the leptoquarks $X$ and $\bar{X}$ do not.

The dynamical superpotential in this model has the same form as the 
one-generation model:
\beq[WdynThree]
\bal
W_{\rm dyn} &= L \bar{\Phi}_{U} \bar{E}
+ L \bar{\Phi}_{D} \bar{N}
+ L Q \bar{X}
+ X \bar{U} \bar{E}
+ X \bar{D} \bar{N}
+ X A \bar{X} + X T \bar{X}
\\
& \quad - \hbox{determinant},
\eal\eeq
where ``determinant'' denotes terms proportional to 9 powers of 
composite meson fields.

% ----------------------------------------------------------------------------
\section{Phenomenological Implications}
% ----------------------------------------------------------------------------
We now discuss some possible phenomenological implications of the 
three-generation model.
The model is far from realistic as it stands:
\susy is not broken, and there are exact horizontal symmetries that 
forbid quark and lepton masses.

From the dynamical superpotential in \Eq{WdynThree}, we see that if 
$\avg{\bar{\Phi}_{U,D}} \ne 0$, then the leptons would become massive.
However, since $\avg{\bar{\Phi}_{U,D}}$ is presumably of order the weak 
scale, this would give equal masses to all leptons of order the weak 
scale.
If $\avg{T} \ne 0$, then the leptoquarks become massive, and lepton 
masses can induce quark masses through loop effects.
However, this would result in up-type quarks that are proportional to 
charged lepton masses, with proportionality $\lsim 1 / (16\pi^{2})$, 
which is clearly unrealistic.

To make the model more realistic, one must add additional fields and 
interactions to break \susy, \ew symmetry, and the horizontal symmetries.
We will not attempt to construct a fully realistic model here, but we 
make some brief comments possible extensions.

% ----------------------------------------------------------------------------
\subsection{Direct Bounds on Compositeness Scale}
We now discuss the phenomenological bounds on the scale $\La$ where 
the hypercolor interactions become strong.
Bounds on flavor-conserving four-fermion interactions give a bound
\cite{PDG}
\beq
\La \gsim \hbox{few}\TeV.
\eeq
The bounds on \FCNCs can be more stringent, but their interpretation
generally depends on the structure of the flavor symmetry breaking.
In the present model, there are interesting \FCNC constraints arising 
from the fact that the quarks and leptons transform under the same 
flavor groups.
In the effective theory below the scale $\La$, this allows 
interactions such as
\beq[Kex]
\de\scr{L} = \frac{1}{\La^{2}} \myint d^{2}\th\, d^{2}\th^{\dagger}\,
\left[ L^{\dagger j} Q^{\dagger}_{j} Q^{k} L_{k}
+ \bar{E}^{\dagger}_{j} \bar{U}^{\dagger j} \bar{U}_{k} \bar{E}^{k}
+ \cdots \right],
\eeq
where we show flavor indices explicitly for clarity.
This will give rise to \FCNC processes conserving
$B_{j} - L_{j}$ and $I_{3j}$, where the charges are nonzero only on 
generation $j$.
The most stringent bound comes from the process
$D^{0} \to e^{-} \mu^{+}$,
which gives a bound
\beq
\La \gsim 4\TeV \left( \frac{f_{D}}{100\MeV} \right)^{1/2}
\left( \frac{m_{c}}{1\GeV} \right)^{-1/2}.
\eeq
There is also a bound from $\pi^{\pm} \to e^{\pm} \nu_{e}$ that comes 
from demanding consistency with $\pi^{\pm} \to \mu^{\pm} \nu_{\mu}$,
which has no non-standard contribution in this model.
This gives a bound $\La \gsim 2\TeV$.
Note that the highly sensitive process $K^{0} \to e^{\pm} \mu^{\mp}$
is forbidden in this model.

% ----------------------------------------------------------------------------
\subsection{Flavor Symmetries}
Because this model naturally has horizontal symmetries, it is 
attractive to consider the possibility that these are spontaneously 
broken.
If the horizontal symmetries are gauged, bounds on flavor-changing
neutral currents imply that the scale of 
spontaneous flavor breaking is larger than $\sim 10^{6}\GeV / g$ if the 
horizontal symmetries are gauged with gauge coupling $g$
\cite{CahnHarari}.
If the horizontal symmetries are global symmetries,
constraints from big-bang nucleosynthesis on 
horizontal Nambu--Goldstone bosons imply that the flavor-breaking 
scale is larger than $\sim 10^{10}\GeV$.
Since we have been unable to find this bound in the literature, we 
explain it briefly.
In order to sufficiently dilute the $3 \times 8 = 24$
Nambu--Goldstone bosons that occur when the flavor symmetry is
broken, they must decouple above the electoweak phase transition.
(The dilution factor is $\sim 1 / 50$ for the minimal 
supersymmetric standard model.)
If we assume that the observed pattern of fermion masses is due to a 
sequential breakdown of the flavor symmetries
$\SU{3} \to \SU{2} \to 1$, then 10 of the Nambu--Goldstone
bosons will be associated with the flavor breaking that gives rise to 
the order-1 top quark Yukawa coupling, and will have couplings
$\sim 1 / f$, where $f$ is the flavor breaking scale.
Demanding that these decouple above $T \sim 100\GeV$ gives the quoted 
bound.

To generate lepton masses, we add to the preonic theory the 
interaction terms
\beq[NewInt]
\de W = \Phi_{U} P_{Q} \bar{P}_{U}
+ \Phi_{U} \De_{U} H_{D}
+ \Phi_{D} P_{Q} \bar{P}_{D}
+ \Phi_{D} \De_{D} H_{U},
\eeq
where we have introduced new elementary fields
\beq\bal
\Phi_{U} &\sim (1, 1, \Yfund, 1) \times
(\bar{\Yfund}, \Yfund, 1, 0, 0),
\\
\Phi_{D} &\sim (1, 1, \Yfund, -1) \times
(\bar{\Yfund}, 1, \Yfund, 0, 0),
\\
\De_{U} &\sim (1, 1, 1, 0) \times
(\Yfund, \bar{\Yfund}, 1, 0, 0),
\\
\De_{D} &\sim (1, 1, 1, 0) \times
(\Yfund, 1, \bar{\Yfund}, 0, 0),
\\
H_{U} &\sim (1, 1, \Yfund, 1) \times
(1, 1, 1, 0, 0).
\\
H_{D} &\sim (1, 1, \Yfund, -1) \times
(1, 1, 1, 0, 0).
\eal\eeq
That is, $\Phi_{U,D}$ are flavored Higgs doublets with quantum 
numbers conjugate to the composite $\bar{\Phi}_{U,D}$ fields,
$H_{U,D}$ are elementary Higgs doublets,
and $\De_{U,D}$ are fields that transform only under the horizontal 
symmetries.

Below the compositeness scale, the first term in \Eq{NewInt} gives
rise to the effective superpotential term
\beq
\de W_{\rm eff} = \La \bar{\Phi}_{U} \Phi_{U}
+ \La \bar{\Phi}_{D} \Phi_{D},
\eeq
so we can integrate out the fields $\Phi_{U,D}$ and $\bar{\Phi}_{U,D}$.
When we do this, the other terms in \Eq{NewInt} give rise to
\beq
\de W_{\rm eff} = \frac{1}{\La} \De_{U} L H_{D} \bar{E}
+ \frac{1}{\La} \De_{D} L H_{U} \bar{N}.
\eeq
Now we assume that there are additional interactions of
$\De_{U,D}$ that give $\avg{\De_{U,D}} \ne 0$, thereby breaking the
horizontal symmetries.
This will give masses to the charged leptons of order
\beq
m_{\ell} \sim \frac{v f_{\ell}}{\La}
\eeq
where $f_{\ell}$ is the \vev that breaks the flavor symmetry of the 
lepton of type $\ell$.
Note that we have introduced no dimensionful couplings into the 
theory.
Also, if $f_{\ell} \ll \La$, this can explain why the charged leptons are
lighter than the weak scale.

If we attempt to extend this mechanism to quarks, we immediately 
encounter a problem.
The fact that the quarks and leptons transform under the same flavor 
symmetry apparently leads to the prediction that the charged lepton
mass matrix is proportional to the up-type quark mass matrix,
which is not even approximately correct.
This conclusion may be avoided by noting that below the compositeness 
scale, the theory has a larger accidental global flavor symmetry under
which the quarks and leptons transform separately.
(In the effective theory below the scale $\La$, this symmetry is 
broken by terms in the effective \Kahler potential such as 
\Eq{Kex}.)
Therefore, flavor symmetry breaking for quarks and leptons can be 
independent.

For this to work, the compositeness scale must be larger than the
scale of flavor symmetry breaking, so we have \eg the bound
$\La \gsim 10^{6} \GeV / g$ for gauged flavor symmetries.
We believe that there are no insurmountable obstacles to 
constructing an explicit model along these lines,
but we will not attempt this here.

% ----------------------------------------------------------------------------
\subsection{Baryon Number Violation}
One very attractive feature of this model is that baryon number 
violation is highly suppressed by accidental symmetries.
Constraints on $B$ violation are very severe:
even dimension-5 operators suppressed by $\sim 1/M\rsub{Planck}$ give
rise to unacceptably large $B$ violation.
The lowest-dimension baryon number violating terms in the preonic
theory are dimension-7 operators of the form
\beq
\de W = \frac{1}{M^{3}}
(\bar{P}_{U} P_{C}) (\bar{P}_{D} P_{C}) (\bar{P}_{D} P_{C}).
\eeq
Since lepton number is not violated, this does not lead to protron 
decay.
The absence of observed neutron--antineutron oscillations gives a
bound $M/\La \gsim 50$.

We will not analyze the precise constraints on higher-dimension 
operators, since it is clear that the theory is safe from Planck-scale 
baryon number violation.
This feature is not spoiled by additional interactions 
needed to make the model realistic as long as these do not involve 
fields that carry baryon or lepton number.

% ----------------------------------------------------------------------------
\subsection{Supersymmetry and Electroweak Symmetry Breaking}
This model does not break \susy as it stands.
If we take the compositeness scale to be large (as suggested by 
considerations of flavor symmetry breaking above), then the usual 
mechanisms of \susy breaking can be used.
That is, \susy may be broken in a hidden sector and communicated to 
the observable sector through gauge, gravitational, or other 
interactions.
(For an alternative approach, see \Ref{LowSUSY}.)

% ----------------------------------------------------------------------------
\section{Conclusions}
% ----------------------------------------------------------------------------
We have constructed a simple class of models that give rise to composite 
quarks, leptons, and Higgs bosons.
The models are based on \susc QCD, with a subgroup of the global 
symmetriesgauged with the standard model gauge group.
Composite fields with quantum numbers of quarks and leptons emerge in 
an interesting way in these models;
for example, it is amusing that leptons are
``baryons'' from the point of view of the hypercolor dynamics.
The models require additional interactions to break \susy, \ew
symmetry, and generate fermion masses.
We hope that this type of model will prove useful in constructing 
realistic composite models of quarks and leptons.

\section{Acknowledgments}
The work of R.N.M. is supported by NSF grant number PHY-9421385.

% ----------------------------------------------------------------------------
% References
% ----------------------------------------------------------------------------

\end{document}